\documentclass[twocolumn,showpacs,aps,prb,superscriptaddress,floatfix,unsortedadresss]{revtex4-1}
\usepackage{graphicx}% Include figure files
\usepackage{dcolumn}% Align table columns on decimal point
\usepackage{bm}% bold math
\usepackage[utf8x]{inputenc}

\begin{document}
 
%
%
%\draft

%\affiliation{Department of Physics and Astronomy, McMaster University, Hamilton, Ontario, L8S 4M1, Canada}
%\affiliation{Department of Physics and Astronomy, Johns Hopkins University, Baltimore, Maryland 21218, USA}
%\affiliation{Canadian Neutron Beam Centre, NRC, Chalk River Laboratories, Chalk River, Ontario, K0J 1J0, Canada}
%\affiliation{National Institute of Standards and Technology, Gaithersburg, Maryland 20899-6102, USA}
%\affiliation{Department of Physics and Astronomy, University of Waterloo, Waterloo, Ontario, N2L 3G1, Canada}
%\affiliation{Brockhouse Institute for Materials Research, McMaster University, Hamilton, Ontario, L8S 4M1, Canada}
%\affiliation{Canadian Institute for Advanced Research, 180 Dundas St. W., Toronto, Ontario, M5G 1Z8, Canada}

\author{J.J. Wagman}
\affiliation{Department of Physics and Astronomy, McMaster University, Hamilton, Ontario, L8S 4M1, Canada}
\author{G. Van Gastel}
\affiliation{Department of Physics and Astronomy, McMaster University, Hamilton, Ontario, L8S 4M1, Canada}

\author{K.A. Ross}
\affiliation{Department of Physics and Astronomy, McMaster University, Hamilton, Ontario, L8S 4M1, Canada}
\affiliation{Department of Physics and Astronomy, Johns Hopkins University, Baltimore, Maryland 21218, USA}
\affiliation{National Institute of Standards and Technology, Gaithersburg, Maryland 20899-6102, USA}

\author{Z. Yamani}
\affiliation{Canadian Neutron Beam Centre, NRC, Chalk River Laboratories, Chalk River, Ontario, K0J 1J0, Canada}
\author{Y. Zhao}
\affiliation{National Institute of Standards and Technology, Gaithersburg, Maryland 20899-6102, USA}
\affiliation{Department of Materials Sciences and Engineering, University of Maryland, College Park, Maryland 20742, USA}

\author{Y. Qiu}
\affiliation{National Institute of Standards and Technology, Gaithersburg, Maryland 20899-6102, USA}
\affiliation{Department of Materials Sciences and Engineering, University of Maryland, College Park, Maryland 20742, USA}

\author{J.R.D. Copley}
\affiliation{National Institute of Standards and Technology, Gaithersburg, Maryland 20899-6102, USA}
\author{A.B. Kallin}
\affiliation{Department of Physics and Astronomy, McMaster University, Hamilton, Ontario, L8S 4M1, Canada}
\affiliation{Department of Physics and Astronomy, University of Waterloo, Waterloo, Ontario, N2L 3G1, Canada}

\author{E. Mazurek}
\affiliation{Department of Physics and Astronomy, McMaster University, Hamilton, Ontario, L8S 4M1, Canada}

\author{J. P. Carlo}
\affiliation{Department of Physics and Astronomy, McMaster University, Hamilton, Ontario, L8S 4M1, Canada}
\affiliation{Canadian Neutron Beam Centre, NRC, Chalk River Laboratories, Chalk River, Ontario, K0J 1J0, Canada}

\author{H.A. Dabkowska}
\affiliation{Brockhouse Institute for Materials Research, McMaster University, Hamilton, Ontario, L8S 4M1, Canada}

\author{B.D. Gaulin}
\affiliation{Department of Physics and Astronomy, McMaster University, Hamilton, Ontario, L8S 4M1, Canada}
\affiliation{Brockhouse Institute for Materials Research, McMaster University, Hamilton, Ontario, L8S 4M1, Canada}
\affiliation{Canadian Institute for Advanced Research, 180 Dundas St. W., Toronto, Ontario, M5G 1Z8, Canada}

\begin{abstract}

We present neutron scattering measurements on single crystals of lightly doped 
$La_{2-x}Ba_{x}CuO_{4}$, with $0\leq x\leq 0.035$. These reveal the evolution of the magnetism in 
this prototypical doped Mott insulator from a three dimensional (3D) commensurate (C) 
antiferromagnetic ground state, which orders at a relatively high T$_{N}$, to a two dimensional (2D) 
incommensurate (IC) ground state with finite ranged static correlations, which appear below a 
relatively low effective T$_{N}$. At low temperatures, the 2D IC magnetism co-exists with 
the 3D C magnetism for doping concentrations as low as $\sim$ 0.0125. We find no signal of a 3D 
C magnetic ground state by x $\sim$ 0.025, consistent with the upper limit of x $\sim$ 0.02 observed 
in the sister family of doped Mott insulators, $La_{2-x}Sr_{x}CuO_{4}$. The 2D IC ground states 
observed for $0.0125 \leq x \leq 0.035$ are diagonal, and are rotated by 45 degrees within the 
orthorhombic basal plane compared with those previously reported for samples with superconducting 
ground states: $La_{2-x}Ba_{x}CuO_{4}$, with $0.05 \leq x \leq 0.095$. We construct a phase diagram 
based solely on magnetic order parameter measurements, which displays much of the complexity of standard 
high temperature superconductivity phase diagrams discussed in the literature. Analysis of high 
energy-resolution inelastic neutron scattering at moderately low temperatures shows a progressive 
depletion of the very low energy dynamic magnetic susceptibility as x increases from 0.0125 to 0.035.  This low energy, 
dynamic susceptibility falls off with increasing temperature on a scale much higher than the 
effective 2D IC T$_{N}$ appropriate to these materials. Appreciable dynamic 2D IC magnetic 
fluctuations inhabit much of the ``pseudogap" regime of the phase diagram.

\end{abstract}

\title{Two Dimensional Incommensurate and Three Dimensional Commensurate Magnetic Order and Fluctuations in $La_{2-x}Ba_{x}CuO_{4}$}

\maketitle

\section{Introduction}

The 214 family of cuprates, $La_{2-x}Ba_{x}CuO_{4}$, and $La_{2-x}Sr_{x}CuO_{4}$, are among the most 
studied of the high temperature superconductors (HTS) \cite{Kastner_RevModPhys_1998,Birgeneau_JPhysSocJpn_2006,
Fujita_JPhysSocJpn_2012}. 
Most of this work has focussed on $La_{2-x}Sr_{x}CuO_{4}$, which has been available in large, 
pristine single crystal form for some time \cite{Keimer_PRL_1991}. Although $La_{2-x}Ba_{x}CuO_{4}$ was 
the original HTS family to be discovered \cite{Bednorz_ZPhysB_1986}, its study has been greatly 
restricted due to the difficulty of its single crystal growth. These difficulties have now been 
largely overcome for relatively low doping levels: $x \leq 0.15$. While there are many similarities 
between the magnetic and 
superconducting properties of the two 214 families of HTS \cite{Dunsiger_PRB_2008,
Matsuda_PRB_1994}, there are also important differences. For example, a low temperature 
tetragonal phase of $La_{2-x}Ba_{x}CuO_{4}$ exists for 0.05 $\leq$ x $<$ 0.15 
\cite{Boni_PRB_1988, Zhao_PRB_2007, Hucker_PRB_2011}, and superconductivity is almost completely suppressed at x = 0.125, 
a phenomenon which is referred to as ``the 1/8 anomaly" \cite{Moodenbaugh_PRB_1988}.

For HTS, the parent, undoped, compounds are Mott insulators which display three dimensional (3D), 
commensurate (C) antiferromagnetic (AF) ground states \cite{Keimer_PRB_1992,Kofu_PRL_2009}. 
This 3D C AF ground state is remarkably sensitive to the presence of mobile, doped 
holes, and less sensitive to the presence of doped mobile electrons \cite{Armitage_RevModPhys_2010}. 
For hole doping, relevant to Ba in $La_{2-x}Ba_{x}CuO_{4}$, Sr in $La_{2-x}Sr_{x}CuO_{4}$ and oxygen, 
in YBa$_2$Cu$_3$O$_{6+x}$, the 3D C AF ground state is very quickly destroyed \cite{Birgeneau_PhysicaB_1992}. 
This occurs, for example, for 
$ x > \sim$ 0.02 in $La_{2-x}Sr_{x}CuO_{4}$\cite{Matsuda_PRB_2000_2, Matsuda_PRB_2000}. 
Upon further introduction of holes, a superconducting ground state is obtained 
for x $\sim$ 0.05 \cite{Matsuda_PRB_2002, Hayden_PRL_1991}. The superconducting $T_{C}$ increases with increased doping and 
an optimally high 
superconducting T$_{C}$ is achieved near x $\sim$ 0.17 \cite{Hayden_PRL_1996, Khaykovich_PRB_2005, Thurston_PRB_1992}. 

Two dimensional (2D) incommensurate (IC) spin structures and dynamics, exhibited by samples with 
hole-doping concentrations beyond those that destroy the 3D C magnetic order, have been studied in 
several families of HTS. Inelastic neutron scattering studies are consistent with an ``hour glass'' 
dispersion, wherein low energy spin excitations disperse out of IC wavevectors and merge or 
nearly-merge at the C wavevector to form a resonant spin excitation
\cite{Tranquada_Nature_2004, Stock_PRB_2007, Vignolle_NaturePhys_2007, Lipscombe_PRL_2007, Lipscombe_PRL_2009, Dai_PRB_2001, Bulut_PRB_1996,Kivelson_RevModPhys_2003}. 
At higher energies, the excitations disperse out from the C wavevector, before 
turning over near the Brillouin zone boundaries \cite{Tranquada_Nature_2004,Coldea_PRL_2001,Stock_PRB_2005}. Such a picture has been 
shown to be relevant even in the relatively low hole-doping regime of $La_{2-x}Sr_{x}CuO_{4}$, where 
the insulating ground state is characterized by so-called ``diagonal" IC spin order 
\cite{Matsuda_PRL_2008}. 

The resulting phase diagrams for these families of HTS materials have led some observers to conclude 
that magnetism and superconductivity are closely linked, as these ground states are either 
contiguous or almost contiguous to each other. In contrast, others have concluded these ground 
states compete, as each inhabits a different part of the phase diagram. From either perspective it 
is important that the microscopic magnetic properties be characterized and well understood across 
the phase diagram. Even in the underdoped, non-superconducting regime, for $x < 0.05$ in 
LBCO and LSCO, the magnetic phase behavior and properties change quickly with doping. This paper 
seeks to elucidate this evolution of magnetic properties in LBCO, using a variety of neutron 
scattering techniques. We specifically report on the magnetic structure and dynamics of 
$La_{2-x}Ba_{x}CuO_{4}$ for doping levels $0 \leq x \leq 0.035$, and study how the 3D C 
magnetism evolves into 2D IC magnetism. We construct a phase diagram for $La_{2-x}Ba_{x}CuO_{4}$ 
based solely on magnetic neutron scattering order parameter measurements and show that it possesses 
much of the full complexity of conventional HTS phase diagrams based on magnetic and transport 
measurements. Finally, using time-of-flight neutron scattering techniques, we report on low 
energy 2D IC spin dynamics in $La_{2-x}Ba_{x}CuO_{4}$ for x $\leq$ 0.035.  We observe the low energy 
dynamic susceptibility to evolve with temperature on a much higher temperature scale than that given 
by the effective 2D IC T$_{N}$ for any doping, and show that 2D IC dynamic spin 
correlations inhabit much of phase diagram associated with the ``pseudogap" state \cite{Timusk_RepProgPhys_1999}.

We will focus our discussion on the 214 cuprate HTS families. For the low doping levels we are 
considering, $La_{2-x}Ba_{x}CuO_{4}$ and $La_{2-x}Sr_{x}CuO_{4}$ are isostructural 
\cite{Katano_PhysicaC_1993, Braden_ZPhysB_1994}. At temperatures that are high relative to room temperature, 
these crystals display tetragonal crystal structures with space group I4/mmm. As the temperature is 
lowered, they undergo a structural phase transition to an orthorhombic structure with space group 
Bmab. This transition occurs near $\sim$ 308 K for x $\sim$ 0.08 in LBCO \cite{Zhao_PRB_2007}, and increases in 
temperature with decreasing doping. We will be presenting measurements on LBCO for x $\leq$ 0.035 
and T $\leq$ 300 K. As such, our samples are orthorhombic at all temperatures measured. The lattice 
parameters in the orthorhombic basal plane are similar, and this has lead some to treat the 
orthorhombic cell as tetragonal for convenience \cite{Hucker_PRB_2011}. We too shall adopt this 
simplification. Doing so, we label the C AF wavevector as $(\frac{1}{2}, \frac{1}{2}, L)$ and 
the diagonal IC ordering wavevectors, relevant for LBCO with x $<$ 0.05, as having the 
form $(\frac{1}{2} \pm \delta, \frac{1}{2} \pm \delta, L)$ and 
$(\frac{1}{2} \mp \delta, \frac{1}{2} \pm \delta, L)$.

\section{Experimental Details}

High quality single crystals of $La_{2-x}Ba_{x}CuO_{4}$ with x = 0, 0.006, 0.0125, 0.025 and 0.035 
were grown by floating zone image furnace techniques using a four-mirror optical furnace. The growth method has been reported on previously \cite{Fujita_PRB_2004, Dabkowska_Springer_2010}. 
The resulting samples were cylindrical in shape and weighed $\sim$ 7 grams each. 
The crystals were all grown in the same excess oxygen atomosphere resulting in a small oxygen 
off-stoichiometry. This oxygen off-stoichiometry could be estimated by measuring the 3D C AF phase 
transition in undoped $La_2CuO_{4+\delta}$, as T$_N$ is known to be sensitive to the precise value 
of $\delta$ \cite{Keimer_PRB_1992_2}. From a determination that T$_{N}$ $\sim$ 250 K for our $La_2CuO_{4+\delta}$ single 
crystal, we estimate that $\delta \approx$ 0.004. We expect this to be the same for all of our 
$La_{2-x}Ba_{x}CuO_{4+\delta}$ samples as they were grown under similar conditions. Hereafter, 
we will not refer to the oxygen off-stoichiometry in the crystals.

Neutron scattering measurements were performed using both time-of-flight and triple axis 
spectrometers. These measurements were carried out using several different cryostats, allowing access to the 
approximate temperature range 1.5 K to 300 K. Three sets of triple axis neutron measurements 
were performed at two laboratories. All measurements were performed with the horizontal scattering 
plane coincident with the HK0 plane of the crystal. Two of these triple axis 
measurements employed a constant final energy E$_{f}$ = 14.7 meV. The first was a set of measurements at the N5 beamline of the NRU reactor at Chalk River Laboratories, which employed a collimation 
of [open-36'-48'-72'] using the convention of collimation between 
[source-monochromator, monochromator and sample, sample and analyser, analyser and 
detector]. The second was a set of measurements using the HB3 instrument at Oak Ridge National Laboratory, which employed 
[48'-40'-40'-120']. Both sets of measurements employed a pyrolitic graphite filter in the scattered 
beam to suppress harmonic contamination, and both had an approximate energy resolution of $\sim$ 
1 meV. High resolution, elastic scattering measurements were also made at Chalk River using E$_{f}$ = 
5.1 meV and collimation of [open, 12', 12', 72']. Both sets of measurements performed at Chalk River employed a cooled beryllium filter for suppression of higher harmonic incident neutrons.  Time-of-flight neutron scattering measurements 
were performed using the NG4 Disk Chopper Spectrometer (DCS)\cite{Copley_ChemPhys_2003} at the National Institute of 
Standards and Technology (NIST) Center for Neutron Research. All DCS measurements presented here 
were performed using an incident neutron wavelenth of 5 $\AA$, and a corresponding energy resolution 
of $\sim$ 0.09 meV. The measurements at DCS were performed with the HHL plane of the 
crystals coincident with the horizontal plane.

\section{Results and Discussion}

\subsection{Magnetic Order Parameter Measurements}

\begin{figure}
\centering
 \includegraphics[width=0.5\textwidth,height=!]{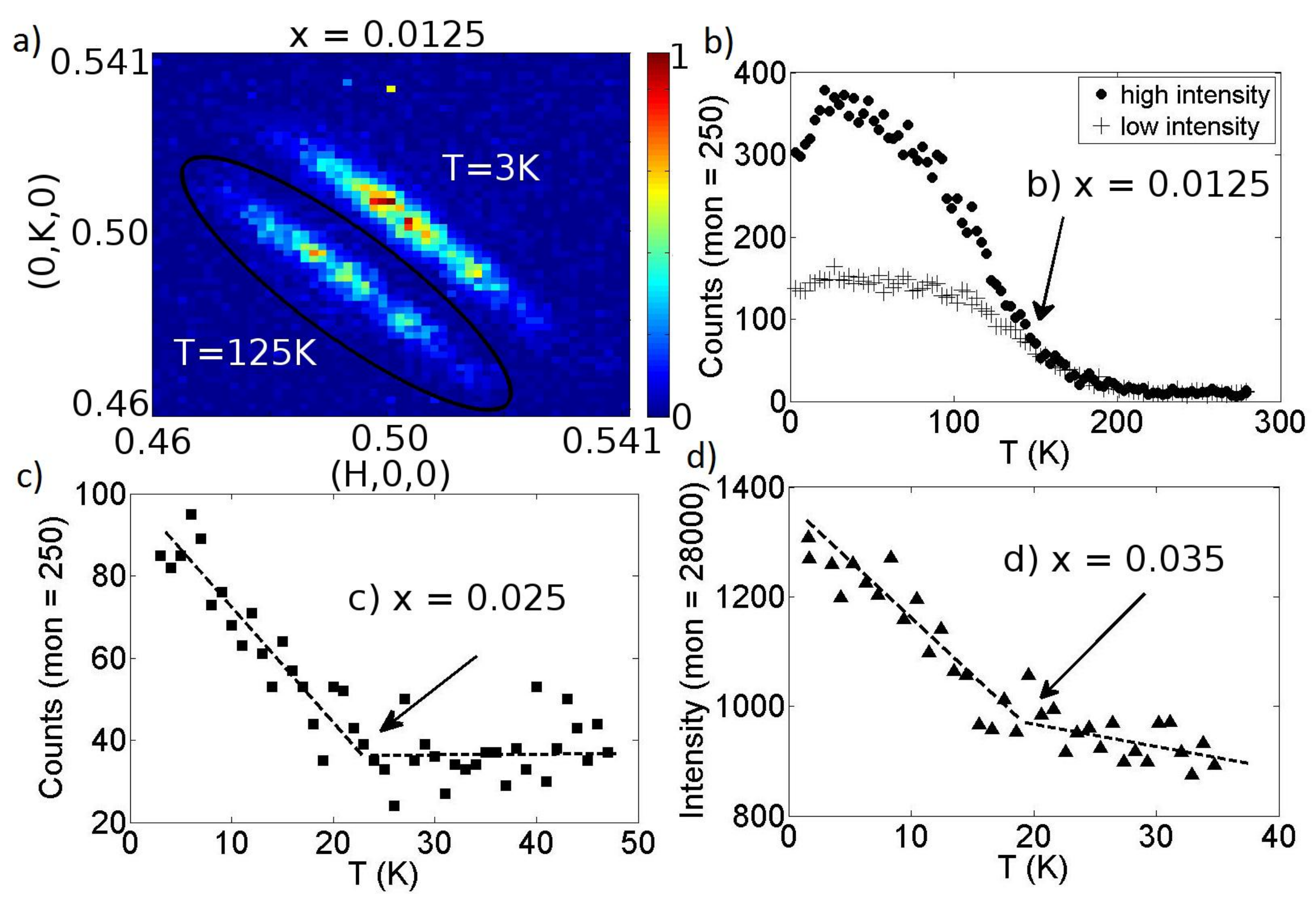} 
\caption{Elastic triple axis neutron scattering measurements of $La_{2-x}Ba_{x}CuO_{4}$ for x = 
0.0125, 0.025 and 0.035 taken on N5 and HB3. a) Reciprocal space maps in the (HK0) plane for x = 0.0125 taken at 3 K and 125 K. The two 
temperature data sets have been shifted for ease of viewing. b) Order parameter measurement for x = 0.0125. The solid circle data are 
measured at the high intensity ``hot spot" identified in the (HK0) map in a). The cross hatched data are collected within the ellipse 
of elastic scattering in a), but away from the ``hot spot". c) and d) Low-resolution, elastic scattering measurements 
of the 2D IC order parameters for x = 0.025 and 
0.035 respectively. Dashed lines serve as guides to the eye.}
 \label{Order_Parameter}
\end{figure}

Magnetic order parameter measurements of 3D Bragg peaks using triple axis spectrometers are 
relatively straight forward to perform, compared with the corresponding measurement of 2D Bragg 
signatures. This is because 3D ordered systems display Bragg spots in reciprocal space, while 
2D Bragg signatures appear as rods in reciprocal space. If the strength of the elastic magnetic 
scattering is otherwise the same, the neutron intensity at a single $\bf{Q}$ position is much larger 
in the 3D case as the signal from the 3D ordered state is localised at a resolution-broadened point 
in reciprocal space, rather than along a rod in the 2D case. Such measurements for a sample of the 
nominally undoped LBCO x = 0 yield a sharp onset to 3D C Bragg scattering at 
$(\frac{1}{2}, \frac{1}{2}, 0)$ for T$_{N}$ = 250 K.  Similar measurements were carried out on 
x = 0.006 and x = 0.0125 samples.

Figure 1 shows the results of such order parameter measurements for x = 0.0125, 0.025, and 0.035 
LBCO samples. Figure 1 also shows high resolution elastic triple axis measurements performed with tight collimations and 
5.1 meV neutrons for the 3D C Bragg peak near $\bf{Q}$ = $(\frac{1}{2}, \frac{1}{2}, 0)$ in the x = 0.0125 sample. Mesh scans in reciprocal space taken at T = 3 K and T = 125 K are shown in 
Fig. 1 a), with the two data sets artificially displaced from each other in the figure for clarity.  In addition,
the intensity scale of each data set has been normalized such that the peak intensity is unity. A broad, elliptical distribution of elastic scattering is observed at all temperatures 
below $\sim$ 200 K. However, as can be seen by comparing the T = 3 K map with the 
intensity-normalized map at T = 125 K in Fig. 1 a), high intensity ``hot spot" develop within 
the ellipse of elastic scattering for temperatures less than $\sim$ 150 K. 

It is possible to follow the temperature dependence of the ``hot spot" scattering and that in the weaker periphery of the ellipse.  Many such measurements were made. The temperature dependence of the sum of all the Bragg scattering at high 
intensity hot spots and at low intensity positions within this ellipse are shown in Fig. 1 b). One can 
see that these two sets of temperature dependencies are the same above $\sim$ 150 K, where both the hot spots and the 
periphery of the Bragg positions show upwards curvature as a function of decreasing temperature. Below $\sim$ 
150 K, the two temperature dependencies markedly depart from each other, with the intensity at the 
hot spot (solid circle data points) positions becoming much stronger than that at the 
corresponding low intensity positions (cross hatched data points). We therefore identify T$_N$ = 150 K for 3D C 
order in our x = 0.0125 single crystal sample of LBCO. 

For the same x = 0.0125 data set at $\sim$ 25 K, we observe a pronounced drop 
off in the intensity of the 3D C AF Bragg scattering. This can be seen in the ``hot spot" order parameter of Fig. 1 b), which 
corresponds to the high intensity positions of the reciprocal space map 
shown in Fig. 1 a). As we will see, this decrease in intensity is associated with the development of 
co-existing 2D IC elastic scattering, which occurs with an ``effective" T$_{N}$ of $\sim$ 25 K for x = 0.0125.

Similar high-resolution elastic magnetic Bragg scattering measurements were performed on 
$La_{2-x}Ba_{x}CuO_{4}$ with x = 0.025 and 0.035, with no obvious sign of 3D C AF order in either 
sample. Lower-resolution elastic scattering measurements were performed using 14.7 meV neutrons 
and relatively coarse collimation, looking explicitly for 2D IC order at appropriate 2D IC diagonal wavevectors 
${\bf Q} = (\frac{1}{2} \pm \delta, \frac{1}{2} \pm \delta, L)$, with $\delta \sim$ x and L = 0, as 
required for measurements within the HK0 scattering plane. The scattering at these 2D IC Bragg positions is weak even at 
low temperatures, as expected for constant-${\bf Q}$ elastic scattering measurements of a 2D rod of 
scattering. Nonetheless effective 2D IC magnetic order parameters were measured and these are shown 
for x = 0.025 and 0.035 samples in Fig. 1 c) and 1 d) respectively. From these measurements we 
identify ``effective" 2D T$_{N}$ of $\sim$ 18 K and 23 K for x = 0.025 and x = 0.035, respectively. 

\subsection{Time-of-Flight Elastic  Neutron Scattering Measurements}

\begin{figure*}[tbp]
\centering 
\includegraphics[width=0.75\textwidth,height=!]{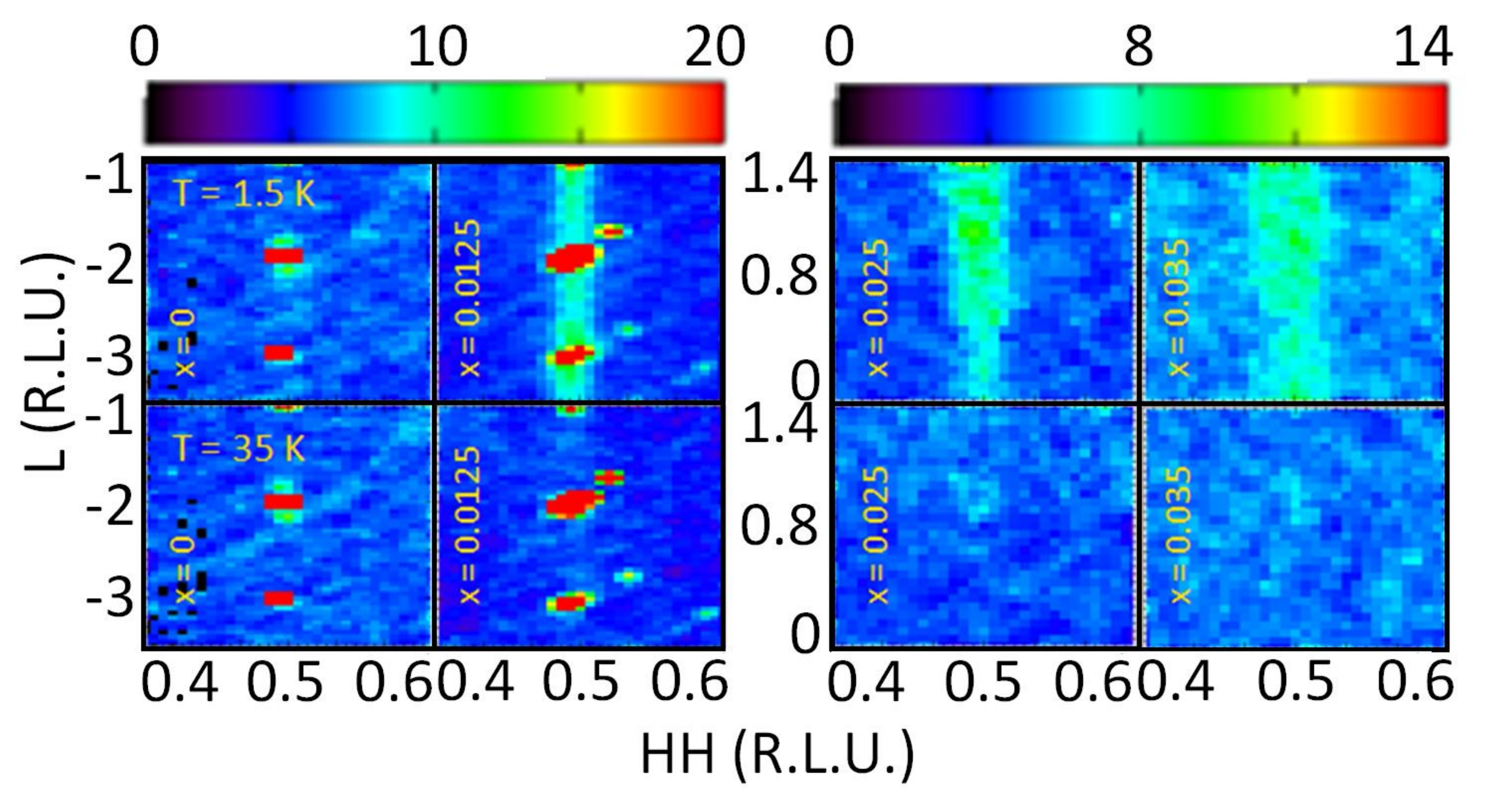} 
\caption{Elastic scattering in $La_{2-x}Ba_{x}CuO_{4}$. From left to right are elastic scattering maps of 
$La_{2-x}Ba_{x}CuO_{4}$ for x = 0, 0.0125, 0.025 and 0.035 respectively. The top row shows
data taken at 1.5 K and the bottom row shows data taken at 35 K. All data have an 
empty cryostat background subtracted from them.}
 \label{Elastic_Maps}
\end{figure*}

The DCS time-of-flight spectrometer was used to measure reciprocal space maps of both the 
elastic, -0.09 meV $\leq \hbar \omega \leq$ 0.09 meV, and inelastic magnetic scattering, 
0.09 meV $\leq$ $\hbar \omega \leq$ $\sim$ 0.8 meV, from our lightly-doped LBCO samples, as a function of temperature. 
Time-of-flight data are shown in Figs. 2, 3, and 4. Figure 2 shows maps of the elastic scattering within the HHL 
scattering plane around $(\frac{1}{2}, \frac{1}{2}, L)$ for four different dopings of $La_{2-x}Ba_{x}CuO_{4}$:
 x = 0, 0.0125, 0.025, and 0.035. The top panels of Fig. 2 shows these maps taken within the ground state of the samples, at T = 1.5 K. 
The bottom panels show the same elastic scattering HHL maps for the same four samples, but now taken at T = 35 K.  This is 
still at a low temperature, but above 25K, which is the ``effective" 2D T$_{N}$ for x = 0.0125 and is the highest for any of these samples. The ranges of L shown were chosen to avoid complications due to absorbtion by the sample.

Three types of Bragg diffraction features can be seen in these reciprocal space maps. Two of these 
features are 3D C Bragg peaks of the form ($\frac{1}{2}$, $\frac{1}{2}$, L = even and L = odd). The 
($\frac{1}{2}$, $\frac{1}{2}$, L = even) 3D C Bragg peaks at L = -2 for x = 0 and x = 0.0125, are 
nuclear-allowed Bragg peaks. The ($\frac{1}{2}$, $\frac{1}{2}$, L = odd) 3D C Bragg peak at L = -3 
for x = 0 and x = 0.0125 is magnetic in origin. Such 3D C magnetic Bragg peaks are absent at all 
temperatures for the x = 0.025 and 0.035 samples, and for the x = 0 and x = 0.0125 above their 
3D T$_{N}$s, $\sim$ 250 and 150 K respectively.

One clearly observes rods of magnetic elastic scattering of the approximate form 
$(\frac{1}{2}, \frac{1}{2}, L)$ for all x except x = 0. These are centred on diagonal IC wavevectors 
$(\frac{1}{2} \pm \delta, \frac{1}{2} \pm \delta, L)$. Note that $\delta \sim$ x is small at these 
low dopings. The rods of scattering are clearly distinct from the Bragg ``spots" which signify 
3D order. Furthermore, these rods show little or no L dependence, a fingerprint of highly-correlated 
2D planes of Cu spin $\frac{1}{2}$ magnetic moments, which are largely decoupled from each other. The only L 
dependence which is observed in our measurements is that associated with self-absorption of the sample in the neutron beam, 
due to the fact that the cylndrical axis of the crystals is not normal to the scattering plane.

Figure 2 shows that the 3D C magnetic order in the x = 0 and 0.0125 samples is largely unaffected by 
raising the temperature from 1.5 K to 35 K. In the x = 0.0125 sample, 2D IC 
static correlations co-exist with 3D C AF order at T = 1.5 K, but no signal of the 2D IC static magnetic 
scattering remains by T = 35 K, leaving only the 3D C AF order. In both the x = 0.025 and 0.035 
samples, only 2D IC static magnetic order exists within the ground state, while the 3D C AF order is 
absent. This is consistent with the low-resolution triple axis measurements on the x = 0.025 and 
0.035 samples shown in Fig. 1 c) and d), wherein the 2D IC magnetism disappears at relatively low 
temperatures, but above their effective 2D T$_{N}$ of $\sim$ 23 K and 18 K, respectively. The appearance 
of the 2D rods of scattering below  $\sim$ 25 K in the x = 0.0125 sample correlates nicely with the suppression of its 3D C 
magnetic Bragg scattering shown in Fig. 1 a).

\begin{figure*}[tbp]
\centering 
\includegraphics[scale=0.125]{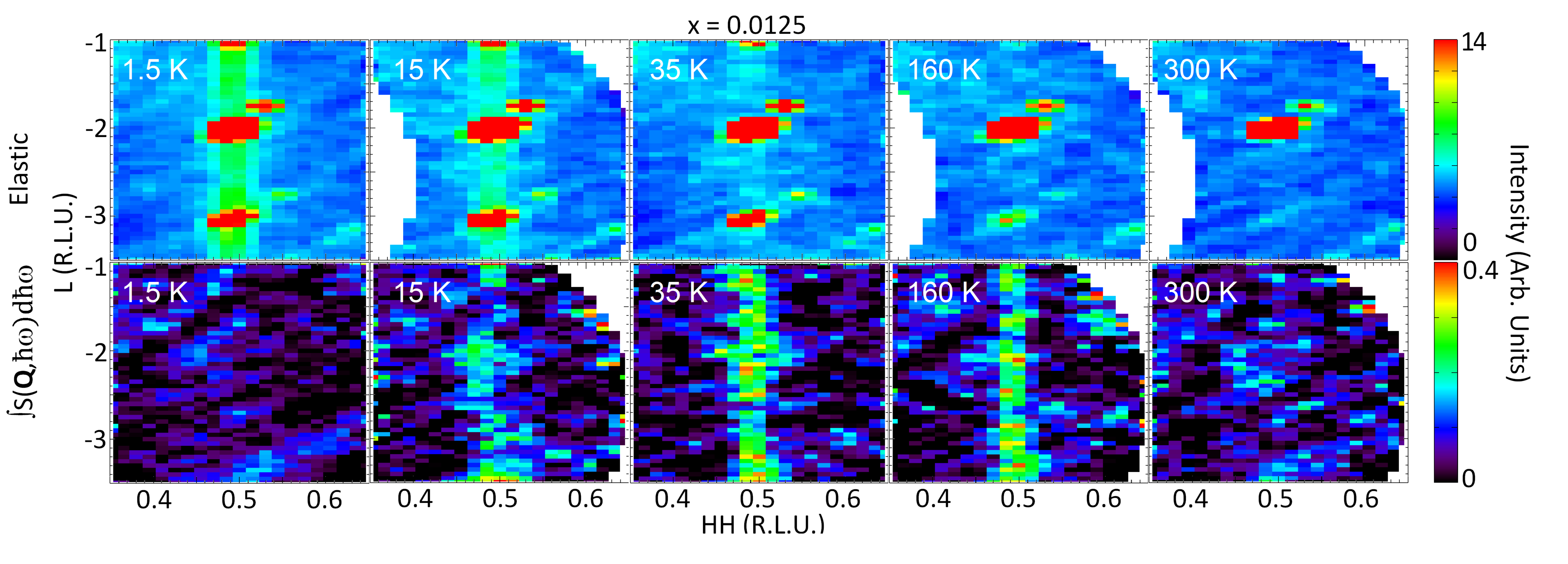} 
\caption{Top row: elastic scattering in $La_{2-x}Ba_{x}CuO_{4}$ for x = 0.0125 shown as a function 
of temperature. Bottom row: Low energy inelastic scattering for the same x = 0.0125 crystal, $S({\bf Q},\hbar\omega)$ integrated between 0.15 meV and 0.8 meV.
Both sets of data were collected in the same time of flight measurement on DCS and 
used the same empty cryostat background subtractions. White areas correspond to regions that were not measured.}
\end{figure*}

The temperature dependence of the magnetic elastic scattering in the x = 0.0125 samples bears 
further attention as both 3D C AF Bragg peaks and 2D IC rods of magnetic scattering coexist within 
the ground state. We have already seen that the temperature dependence of the 3D C AF 
Bragg peak for x = 0.0125, shown in Fig. 1 b), has a reduction in the scattered intensity below 
$\sim$ 25 K. The top set of panels in Figure 3 shows the same $(\frac{1}{2}, \frac{1}{2}, L)$ 
elastic reciprocal space map for x = 0.0125 shown in Fig. 2, now as a function of temperature. For 
now, we will focus only on the top panels and will return to the bottom panels when we discuss the 
inelastic scattering in a later section. We clearly see the disappearance of the rod of elastic scattering as the 
temperature increases to T = 35 K, and that the 3D C AF peak near $(\frac{1}{2}, \frac{1}{2}, -3)$ 
has all but disappeared at T = 160 K, above the 3D C T$_{N} \sim$ 150 K, indentified 
in Fig. 1 b) from high-resolution triple axis order parameter measurements. We note that the nuclear 
Bragg peak at $(\frac{1}{2}, \frac{1}{2}, -2)$ is nearly temperature independent over the range of 
temperature shown. This is as expected for a nuclear Bragg peak, given that all temperatures studied 
are well removed from the orthorhombic-tetragonal structural phase transition in this material\cite{Reehuis_PRB_2006}.

The trade-off between 3D C AF and 2D IC static magnetism shown in Fig. 3 and Fig. 1 b) is similar to 
that reported for $La_{2-x}Sr_{x}CuO_{4}$ at similar doping levels \cite{Birgeneau_JPhysSocJpn_2006}. 
It implies that the 3D C AF structure forms as the temperature is lowered, but that part of this 
structure is unstable to the formation of 2D IC order below the 2D effective T$_{N}$ of 25 K 
in the x = 0.0125 sample. While it is not easy to compare the integrated intensity of 
the 2D rod scattering to the 3D C AF Bragg scattering, it is straightforward to estimate the 
reduction of the 3D C AF Bragg peak from saturation shown below $\sim$ 25 K in Fig. 1 b). This shows 
that for x = 0.0125 the 2D IC static order accounts for 20$\%$ of the elastic magnetic scattering 
in the ground state. As suggested for $La_{2-x}Sr_{x}CuO_{4}$, this fraction presumably grows with 
x until it accounts for $100 \%$ of the static elastic magnetic scattering in the ground state for 
x $\geq$ 0.02 \cite{Matsuda_PRB_2002}.

The HHL reciprocal space maps around $(\frac{1}{2}, \frac{1}{2}, L)$, shown in  Fig. 2, all cover 
the same range in (HH). Note that the L ranges shown differ due to the fact that the self absorbtion 
for a given position in the (HHL) plane differs for the four crystals. It is clear that at T = 1.5 K 
(the top panels in Fig. 2) the rod of magnetic scattering broadens in the (HH) direction 
progressively with increasing doping from x = 0.0125 to 0.035. This is due to the fact that 2D IC 
static order is expected to change its incommensuration with doping, and the expected dependence is 
roughly $\delta \sim$ x in the diagonal IC wavevevctor 
$(\frac{1}{2} \pm \delta, \frac{1}{2} \pm \delta, L)$ 
\cite{Yamada_PRB_1998,Enoki_PRL_2013}. 

\begin{figure}
\includegraphics[width=0.5\textwidth,height=!]{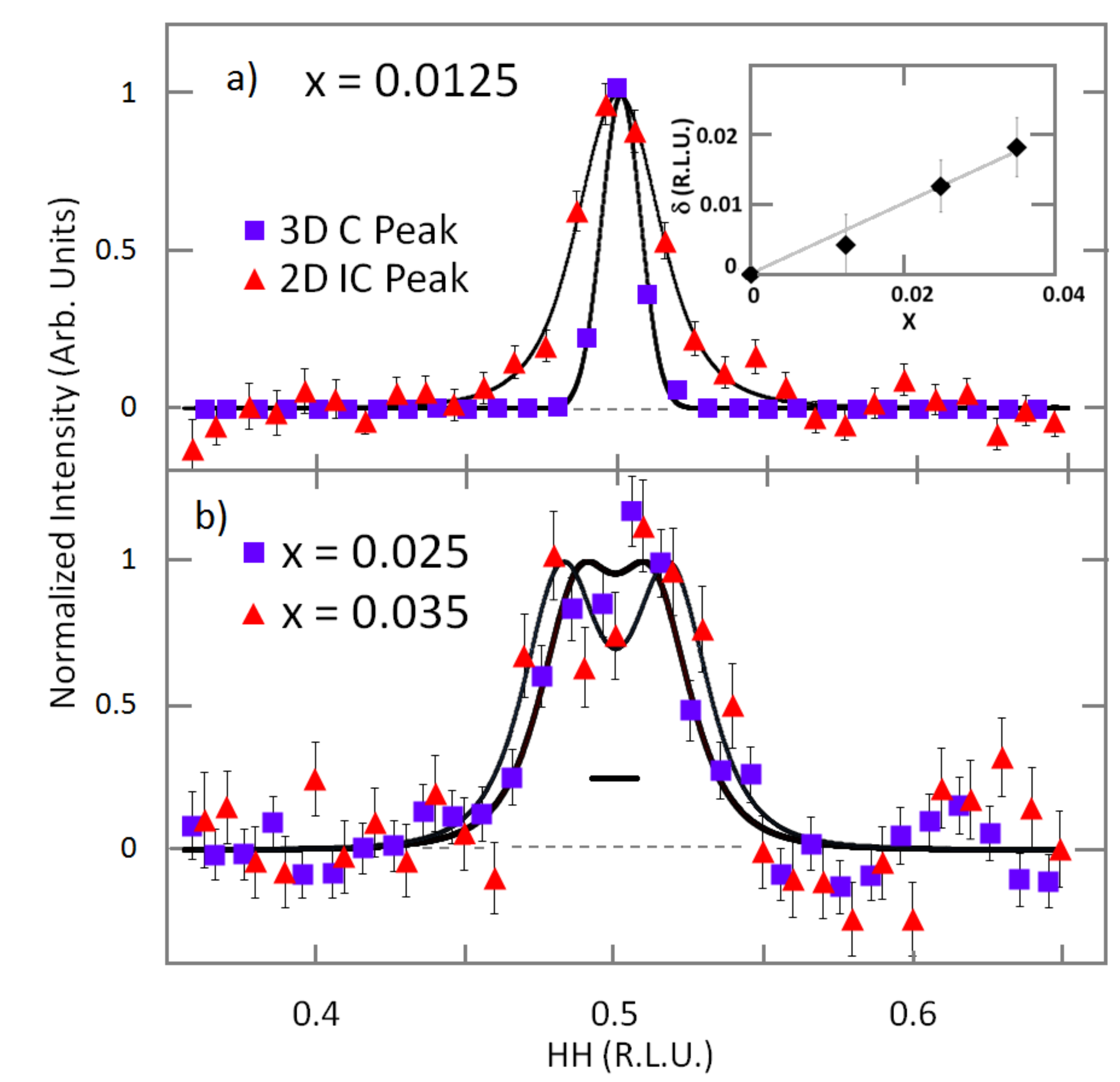} 
\caption{Cuts through the elastic magnetic scattering are shown for x = 0.0125, 0.025, and 0.035. 
Data sets have been normalized to their own maximum 
intensity for the purposes of qualitative comparison. Solid lines are fits to the data as discussed 
in the text. a) 3D C structural and 2D IC magnetic peaks are shown in x = 0.0125 at T = 1.5 K. The 2D IC for this doping 
integrated the data over the [-3.6,-3.2], [-2.7,-2.2],[-1.6,-1.2],[-0.9,-0.65] ranges in L, so as to avoid contributions from 3D C peaks. 
The 3D C structural peak corresponds to L = -4, and employed a $-4.1 \leq L \leq -3.9$ integration. b) 2D IC peaks in 
x = 0.025 and 0.035 using $0 \leq L \leq 1.9$ integration for both samples. This range avoids contributions from nuclear 
Bragg peaks and is minimally affected by self absorbtion. Inset: incommensuration $\delta$ as a function of doping as determined from fits of the data. Error bars represent one standard deviation.}
\end{figure}

We can explicitly examine the lineshape associated with the 2D IC rods of magnetic scattering and 
estimate both the $\delta$ vs x dependence in the ground state, and the finite-range of the in-plane 
spin correlations within the 2D IC structure. In Fig. 4, we show cuts in the (HH) direction through 
the reciprocal space maps displayed in the top row of Fig. 2 appropriate to T = 1.5 K. For the 
x = 0.0125, 0.025, and 0.035 data sets, the cuts are taken so as to pick out only the 2D IC rod 
scattering; that is they sample data between the nuclear allowed 
$(\frac{1}{2}, \frac{1}{2}, L = even)$ 3D Bragg peaks for all samples, as well as avoiding the 
3D C magnetic Bragg peaks at $(\frac{1}{2}, \frac{1}{2}, L = odd)$ for the x = 0.0125 sample. 
For comparison, we also have a cut through the L = -4 structural Bragg peak in the x = 0.0125 sample as 
a measure of the instrumental resolution.

These cuts are shown in Fig. 4 a) and b). The x = 0.0125 3D C data set is clearly much narrower in (HH) than 
that of any of the other three data sets, which exhibit 2D diagonal IC order. The three 2D IC data 
sets were fit phenomenologically to a functional form of two squared-Lorentzians with 
identical widths and amplitudes, but centered at different HH positions. Previous studies of such quasi-two 
dimensional correlations also employed Lorentzian-squared lineshapes to describe the IC elastic 
scattering \cite{Matsuda_PRL_2008}. Initially, these data were fit with the sum of two 
Lorentzians-squared lineshapes wherein their widths were allowed to vary with x. However, the resulting 
variation of the width with x was small, and the fits were redone using a common width for the 
Lorentzian-squared lineshapes in all fits.

As the finite (HH) width to the Lorentzian-squared lineshape represents a finite (inverse) 
correlation length, we conclude that the 2D IC static order is short ranged in 
$La_{2-x}Ba_{x}CuO_{4}$, with a correlation length of $\sim 20 \AA$. 
Over this doping range and to within our resolution, this correlation length is independent of doping. The diagonal 
IC wavevector $\delta$ can then be extracted from this analysis, and this is shown as a function of 
x at T = 1.5 K in the inset to Fig. 4 a). We observe a linear relationship $\delta\sim$ x for x = 0.125, 
0.025, and 0.035, which extrapolates back through zero at x = 0. This conclusion is somewhat 
different from that reached in previous studies of $La_{2-x}Sr_{x}CuO_{4}$, wherein a linear 
$\delta\sim$ x relationship was also found for sufficiently large x, but $\delta$ was $\sim$ independent 
of x for very low concentrations $<$ 0.02, which also displayed 3D C AF order\cite{Wakimoto_PRB_2000, Yamada_PRB_1998}.

\subsection{Time-of-flight Inelastic Scattering Measurements and Dynamic Susceptibility}

The DCS time-of-flight instrument allows the simultaneous measurement of elastic neutron 
scattering and inelastic neutron scattering.  Reciprocal space maps of the inelastic scattering can also be constructed, similar to the elastic 
scattering data presented in Figs. 2 and 3.    The relatively low incident energy, E$_i$, employed in these measurements restricts the accessible inelastic 
scattering to less than $\sim$ 1 meV energy transfer, although the magnetic excitations in this system are known to exist to 
significantly higher energy\cite{Coldea_PRL_2001}. We have plotted inelastic scattering for the x = 0.0125 sample as a function 
of temperature in the bottom panels of Fig. 3. A comparison between this magnetic inelastic 
scattering from the 0.0125, 0,025, and 0.035 samples, all at T = 35 K, is shown in the top panel of 
Fig. 5.

\begin{figure*}[tbp]
\centering 
\includegraphics[scale=0.7]{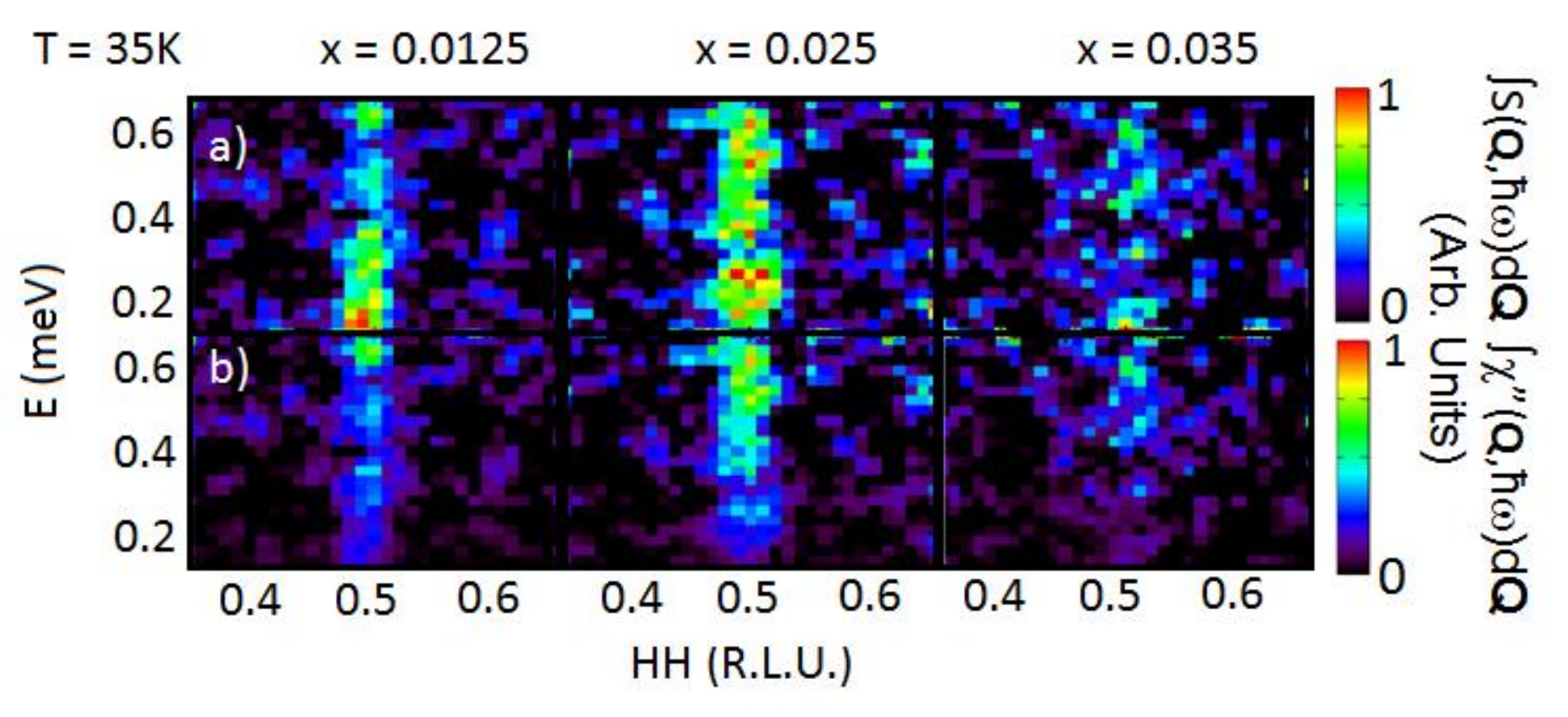} 
\caption{Inelastic scattering for x = 0.0125, 0.025 and 0.035 at T = 35 K.  All data sets employed a T=1.5 K
data set as background.  Panels a) and b) show energy-HH wavevector 
maps for S(${\bf Q}$,$\hbar\omega$) (panel a)) and $\chi^{\prime\prime}$(${\bf Q}$,$\hbar\omega$) (panel b)). These data sets employed 
$-3.5 \leq L \leq -1$ integration for x = 0.0125 and $0 \leq L \leq 1.9$ integrations for x = 0.025 and 0.035.
$\chi^{\prime\prime}$(${\bf Q}$,$\hbar\omega$) is related to S(${\bf Q}$,$\hbar\omega$) through Eqs. 1 and 2.}
\end{figure*}

The bottom panels of Fig. 3 show reciprocal space maps of the inelastic scattering from the 
x = 0.0125 sample, integrated in energy from 0.1 meV to 1 meV, and as a function of temperature 
between T = 1.5 K and T = 300 K. This integrated inelastic magnetic scattering can be compared 
directly to the same reciprocal space maps of the elastic scattering around 
$(\frac{1}{2}, \frac{1}{2}, L)$ wavevevectors as shown in the top panels for Fig. 3. On this relatively low energy 
scale, we observe an interesting trend wherein little inelastic scattering is observed at 
T = 1.5 K, although both the 2D IC elastic rod of magnetic scattering and the 3D C AF Bragg peaks 
are strong. As the elastic rod of 2D IC scattering for the x = 0.0125 sample fades in intensity above 
T = 15 K, the inelastic scattering becomes clearly evident. Above the 2D effective T$_{N} \sim$ 
23 K, only the 2D IC inelastic scattering and the 3D C elastic magnetic scattering remain. 
The intensity of the 2D IC inelastic scattering is prevalent out to 160 K, but has clearly faded 
at the highest temperature measured, T = 300 K. The lower panels of Fig. 3 
show 2D IC dynamic spin fluctuations in the x = 0.0125 sample are present well above the effective 
2D T$_{N} \sim$ 23 K, and only completely disappear above the temperature characteristic of the 3D C T$_{N}, 
\sim$ 150 K, in this sample.

The evolution of the low energy inelastic magnetic scattering and the corresponding imaginary part 
of the dynamic susceptibility as a function of doping is shown in Fig. 5. The top panel of Fig. 5 a) 
shows the inelastic scattering at T = 35 K for each of the x = 0.0125, 0.025 and 0.035 samples. 
There data has been  integrated in L using $-3.5 < L < -1$ for x = 0.0125 and $0 < L < 1.9$ in 
x = 0.025 and 0.035. The reason for this choice of L integration is that these regimes avoid 
complications due to self absorption that arise as the sample is rotated in the beam. This data 
is plotted in an energy vs. (HH) wavevevector map, over the approximate range in energy from 
0.15 meV to 0.8 meV.  A temperature of 35 K was chosen for this comparison as it is sufficiently low 
to approximate the ground state, while high enough such that appreciable magnetic 
inelastic intensity is evident in all samples. We note that there is little magnetic inelastic 
scattering evident at T = 1.5K. We take advantage of this and use the T = 1.5K data sets as a 
measure of the inelastic background for our samples. This will be important in isolating the dynamic 
magnetic susecptibility from our inelastic scattering data.

\begin{figure*}[tbp]
\centering 
\includegraphics[width=1\textwidth,height=!]{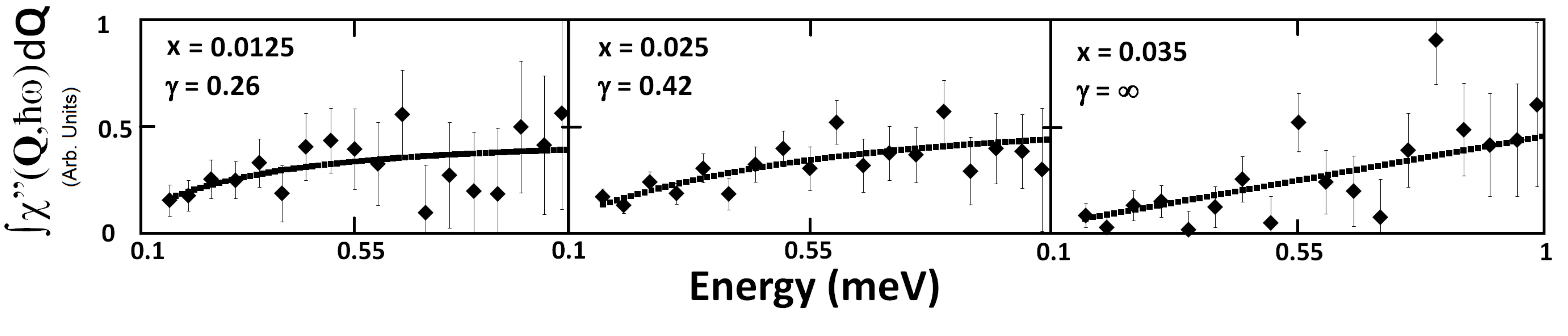} 
\caption{The energy dependence of $\chi^{\prime\prime}$(${\bf Q}$,$\hbar\omega$) integrated over the HH width of the rod of scattering shown in Fig. 5 b), 
as described in the text. The dashed line shows a fit to a phenomenological model, Eq. 3, describing this energy dependence.}
\end{figure*}

The magnetic inelastic scattering, expressed as S(${\bf Q}$, $\hbar\omega$, T) is the product of two terms; the Bose population 
factor which maintains detailed balance, and the imaginary part of the dynamic 
susceptibility $\chi^{\prime\prime}$. $\chi^{\prime\prime}$ is an odd function of energy and 
characterizes the capacity of the system to absorb energy, thereby creating spin excitations at a particular wavevector and 
energy. The inelastic magnetic scattering, S($\bf{Q}$, $\hbar\omega$, T) is then related to 
the imaginary part of the dynamic susceptibility through:

\begin{equation}
 S({\bf Q},\hbar\omega,T) = [n(\hbar\omega,T)+1] \times \chi^{\prime\prime}({\bf Q},\hbar\omega,T)
\end{equation}

where $[n(\hbar\omega,T)+1]$ is the Bose population factor:

\begin{equation}
 [n(\hbar\omega,T)+1] = \frac{1}{1-e^{\frac{-\hbar\omega}{k_{B}T}}}
\end{equation}

At T = 35 K and for energies $\leq$ 1 meV, the Bose population factor, $[n(\hbar\omega,T)+1]$, 
is sufficiently strong that the overall neutron scattering signal, S(${\bf Q}$, $\hbar\omega$) can be easily 
distinguished from background for all concentrations. One can isolate S(${\bf Q}$, $\hbar\omega$) with an appropriate subtraction and divide 
through by the Bose factor to give the imaginary part of the dynamic susceptibility, 
$\chi^{\prime\prime}$. This is what is shown in the b) panels of Fig. 5 for x = 0.0125, 0.025, and 0.035, 
listed from left to right. The a) panels of Fig. 5 show the corresponding S(${\bf Q}$, $\hbar\omega$). 
Focussing on $\chi^{\prime\prime}({\bf Q},\hbar\omega)$ in the b) panels of Fig. 5, we see a 
suppression of $\chi^{\prime\prime}({\bf Q},\hbar\omega)$ at low energies. This suppression 
increases with doping between 0.0125 and 0.035. In Fig. 6 we show cuts of 
$\chi^{\prime\prime}({\bf Q},\hbar\omega)$ made by integrating the data in Fig. 5 b) in (HH) 
around $0.48 \leq$ (HH) $\leq 0.52$ and over the relevant L-range so as to capture all the dynamic 
magnetic susceptibility in this low energy regime. The resulting quantity is then plotted in 
Fig. 6 as a function of energy for x = 0.0125, 0.025, and 0.035. We see a suppression of the low energy dynamic susceptibility as the doping increases. 
This can be quantified by fitting the energy dependence of this integrated low energy dynamic 
susceptibility to the phenomenological form\cite{Aeppli_DPUMS_1998}:

\begin{equation}
 \chi^{\prime\prime}(E) = A \times tan^{-1}(E/\gamma)
\end{equation}

This allows the extraction of a characteristic energy scale, $\gamma$, at which the magnetic dynamic 
susceptibility, as a function of decreasing energy, turns over and decreases towards zero, as it must in order to be an odd function of 
energy. The fit is displayed as the dashed lines in Fig. 6, and the appropriate $\gamma$ value 
resulting from the fit is displayed in the left corner of each panel.
 
As expected from Fig. 5 b), $\gamma$ is lowest for the x = 0.0125 sample, and increases with 
increased hole doping, x. It is interesting to note that this progression is established in samples that are not superconducting. 
One might expect this phenomena to be linked to the superconducting gap that is observed in LSCO. There, it is known that for samples of LSCO with superconducting ground states, that is 
$x \geq 0.05$, a spin gap forms for T $< T_{C}$ within the dynamic susceptibility at low energies. For example, for samples with x = 0.16 it is reported that the gap is 7 meV\cite{Christensen_PRL_2004}. 
That said, it has also been reported that no such corresponding spin gap exists 
in LBCO out to at least x = 1/8 \cite{Fujita_PRB_2004}. This has been motivated by the absence of a temperature dependence to the low
energy dynamic susceptibility in underdoped superconducting samples.  However, there has been suggestion that a superconducting gap may exist 
for higher dopings \cite{Hucker_PRB_2011}. To be sure, the present spin gap related phenomena may be different from the related phenomena which occurs in LSCO. 
But, it is clear that an interesting depletion in dynamic susecptibility as a function of increasing doping seems to be a characteristic of LBCO as well.

\begin{figure*}[tbp]
\centering 
\includegraphics[scale=0.15]{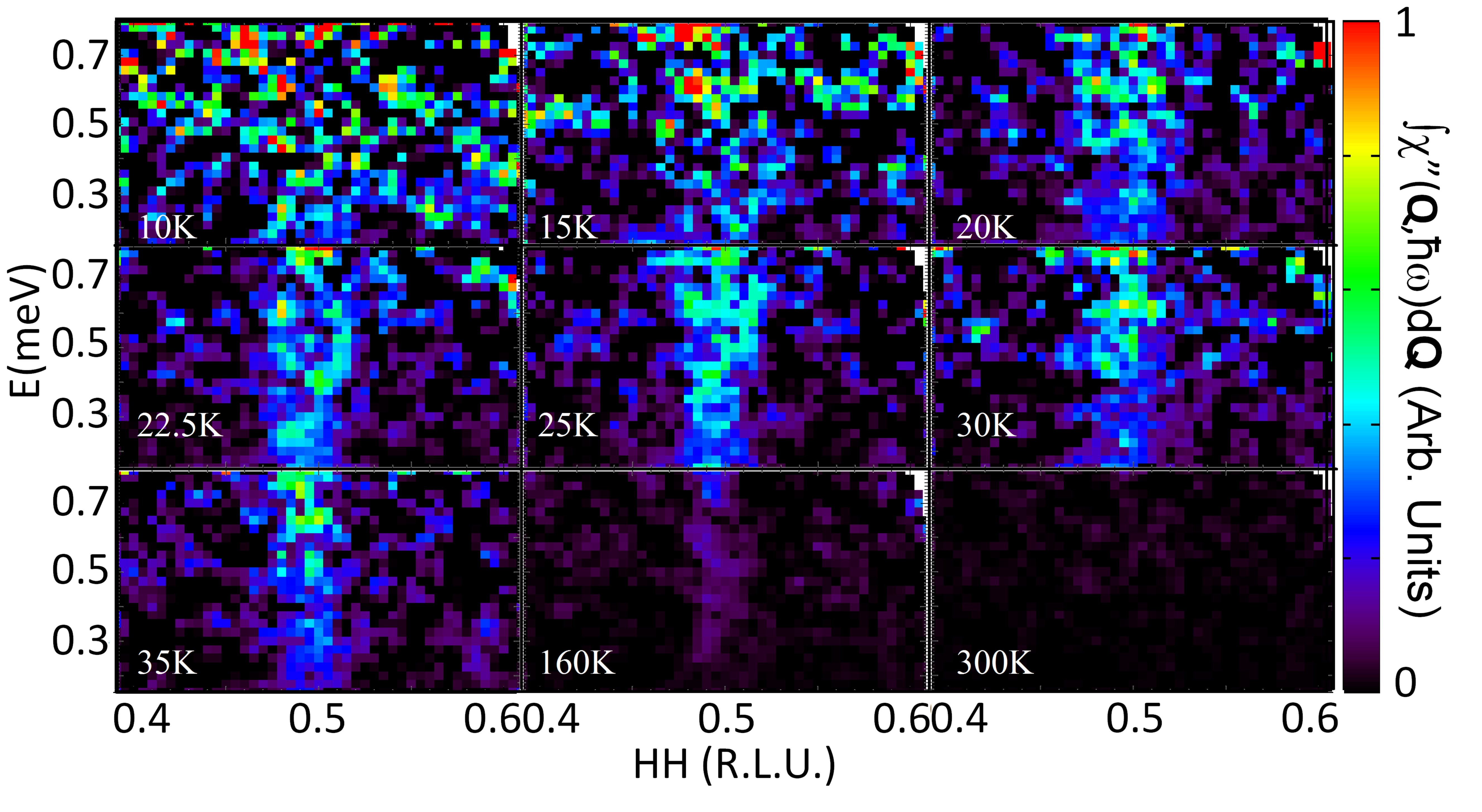} 
\caption{Energy-wavevector maps of $\chi"({\bf Q},\hbar\omega)$ for x = 0.0125 are shown as a function of temperature, from 
10 K to 300 K. There is clear monotonic decrease in the spectral weight of the dynamic magnetism with tempertaure.}  
\end{figure*}

\begin{figure*}[tbp]
\centering 
\includegraphics[scale=0.15]{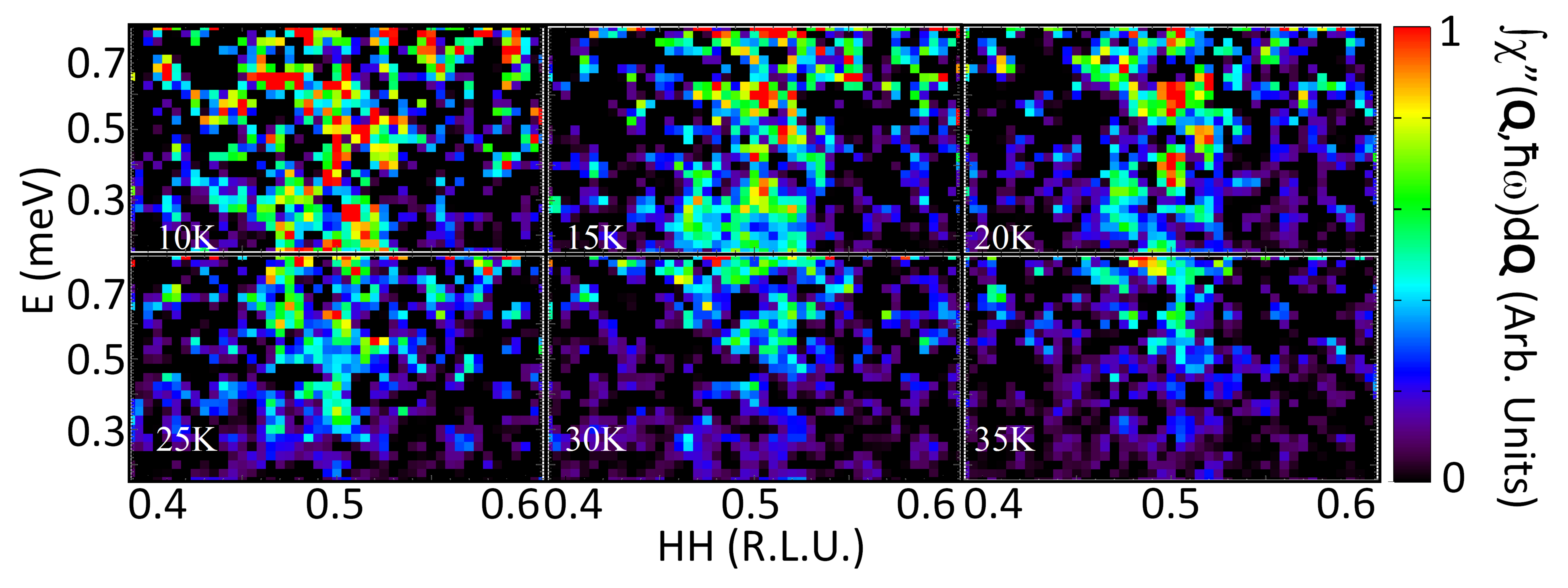} 
\caption{Energy-wavevector maps of $\chi"({\bf Q},\hbar\omega)$ for x = 0.035 are shown as a function of temperature, from 10 K to 35 K.}
\end{figure*}

We now examine the temperature dependence of $\chi^{\prime\prime}({\bf Q},\hbar\omega)$ for 
x = 0.0125 and 0.035 samples in Figs. 7, 8, and 9. Figure 7 shows 
$\chi^{\prime\prime}({\bf Q},\hbar\omega)$ for x = 0.0125 in energy vs. (HH) maps over the range of 
energy from 0.15 to 0.8 meV, again integrated in L around the ranges appropriate to isolate 2D rods of scattering, as used in Fig. 5. 
These data sets are at temperatures ranging from 10 K to 300 K, as denoted in the bottom left of 
each panel, and all data sets used T = 1.5 K data sets as background. Figure 8 shows the same 
$\chi^{\prime\prime}({\bf Q},\hbar\omega)$ maps for the x = 0.035 sample over the temperature range 
T = 10 K to T = 35 K, again using the appropriate T = 1.5 K data set as a background.

In both the x = 0.0125 and 0.035 cases, $\chi^{\prime\prime}({\bf Q},\hbar\omega)$ clearly decreases 
monotonically with increasing temperature over this relatively low energy range. The intriguing behavior seen in 
the bottom panels of Fig. 3 for the x = 0.0125 sample, wherein the 2D IC inelastic intensity appears 
to have a temperature dependence complementary to that of the 2D IC elastic scattering, can be 
understood as a consequence of the temperature dependence of the Bose factor, $[n(\omega)+1]$.

To better understand $\chi^{\prime\prime}({\bf Q},\hbar\omega, T)$ quantitatively, we integrated the 
$\chi^{\prime\prime}({\bf Q},\hbar\omega)$ data shown in Figs. 7 and 8 in energy between 0.2 and 0.8 
meV. This was then fit to a Gaussian lineshape centred on HH=$(\frac{1}{2}, \frac{1}{2})$ with a linear background.
The integrated intensity of the Gaussian gives $\chi^{\prime\prime}$(${\bf Q}$ $\sim$ $(\frac{1}{2}, \frac{1}{2}, L),0.2 meV \leq \hbar\omega \leq 0.8 meV)$,
which is plotted as a functon of temperature on a semi-log scale in Fig. 9 for both x = 0.0125 
and 0.035 samples. The signals from the x = 0.0125 and 0.035 samples have been approximately normalized at low temperatures. 
The values of T$_{N(2D IC)}$ for x = 0.0125 (25K) and 
0.035 (15K) as well as T$_{N(3D C)}$ for x = 0.0125 (150K) and x = 0 (250K) are indicated for reference as 
dashed lines in Fig. 9. We find that the dynamic IC magnetism in both samples is present on a temperature 
scale that is independent of the static ordering temperatures in either system. $\chi^{\prime\prime}({\bf Q},\hbar\omega)$ is strongest 
at low temperatures in both materials, and its temperature dependence does not suggest a well defined transition temperature. The 
phenomenon observed is instead consistent with a cross-over that occurs at some temperature above the 3D C T$_N$ for the x = 0.0125 system. 

\begin{figure}
\includegraphics[width=0.5\textwidth,height=!]{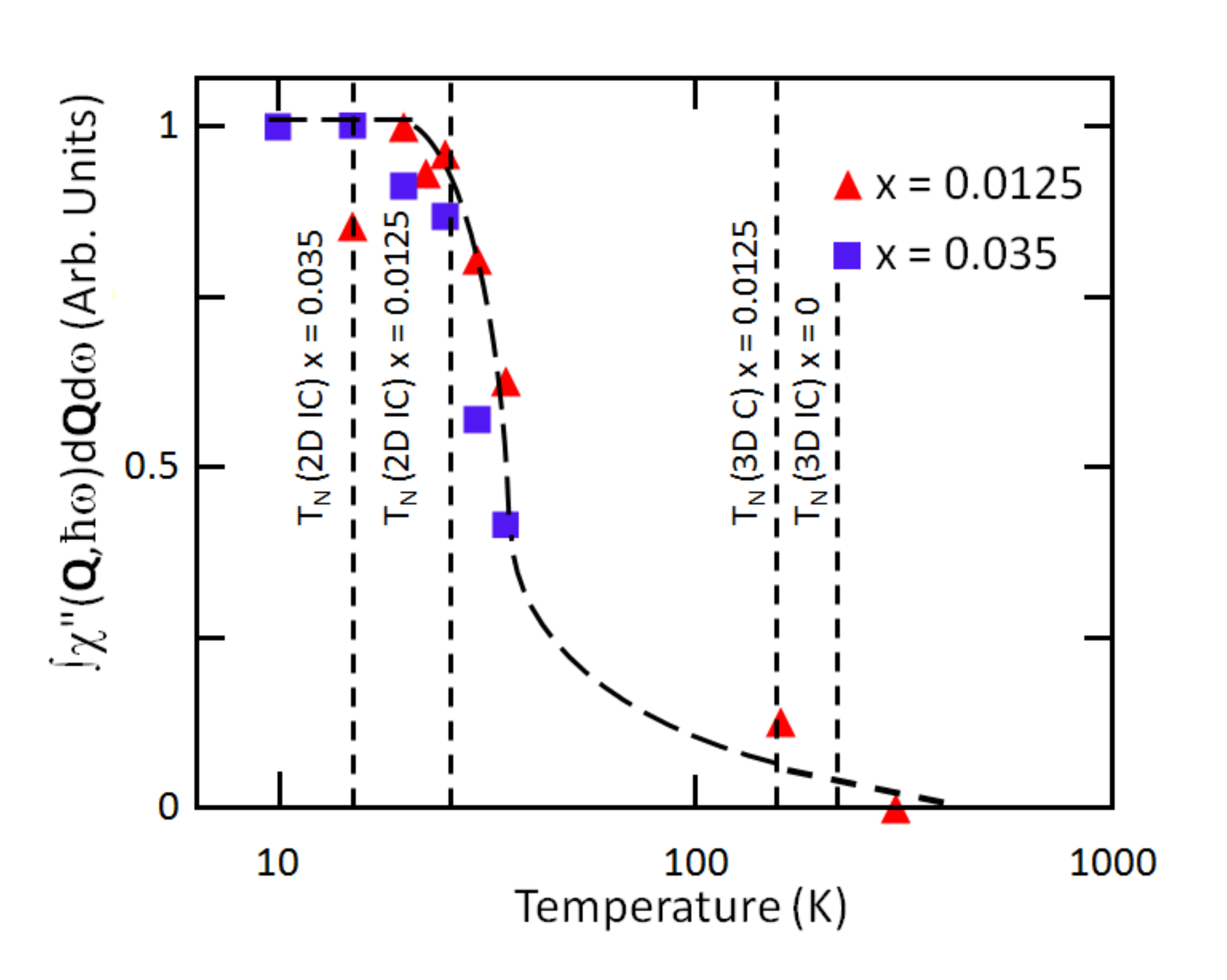} 
\caption{Temperature dependence of the wavevector and low energy (0.2 meV $\leq$ $\hbar\omega$ $\leq$ 0.8 meV) integrated $\chi^{\prime\prime}({{\bf Q},\hbar\omega})$ for x = 0.0125 and 0.035
is shown as a function of temperature on a semi-log scale.  Dashed lines show the 2D IC (for x = 0.035, 0.0125) and 
3D C (for x = 0.0125 and 0) magnetic ordering temperatures. The x = 0.0125 and 0.035 data sets have been normalized at low temperatures.  
Both data sets have employed their 1.5K data set as a background. The temperature scale for the evolution of this low energy dynamic magnetism greatly exceeds the relevant 
2D IC magnetic ordering temperatures.}
\end{figure}

\subsection{Magnetic Phase Diagram}

We summarize our elastic and inelastic magnetic neutron scattering measurements on relatively lightly-doped 
$La_{2-x}Ba_{x}CuO_{4}$ in the phase diagram shown in Fig. 10. It displays three sets of points 
which represent phase transition temperatures appropriate to 3D C AF order (red circles), 2D 
diagonal IC static order (yellow triangles), and 2D parallel IC order (blue circles). The latter set 
of phase transitions occur for concentrations with superconducting ground states for x $\geq$ 0.05, 
and comes from our earlier neutron results on magnetic order parameter measurements\cite{Dunsiger_PRB_2008,Dunsiger_PRB_2008_2}.  

We also show extended regions on the phase diagram where 2D dynamic IC magnetism is observed. As seen 
in Fig. 9, this dynamic 2D IC magnetism gradually fades with increasing temperature and does not 
display an obvious phase transition. This dynamic 2D IC magnetism occupies the same general region 
of the HTS phase diagram associated with the ``pseudogap phase''. The pseudogap phase has been 
ascribed to several different origins, including phase-incoherent superconducting pairs 
\cite{Emery_Nature_1995}, and ordering associated with orbital currents \cite{Varma_PRB_1997}. 
Whatever other properties it possesses, it is clear that 2D IC spin fluctuations are strong 
throughout this entire region and that the crossover to a fully paramagnetic state occurs on a high 
temperature scale.

Coming back to the 3D C and 2D IC phase transitions identified from elastic neutron scattering order 
parameter measurements, shown as the circles, triangles, and squares respectively in Fig. 10, there 
are several interesting observations to make. First and foremost, Fig. 10 is compiled exclusively 
from magnetic order parameter measurements. Nevertheless, it displays much of the full complexity of the HTS 
phase diagram. In our opinion, such an observation in and of itself leads to the conclusion that the 
superconducting ground state is intimately related to the magnetic ground state.  Second, 
the fact that the 2D effective T$_{N}$ is so much smaller than the 3D C T$_{N}$ is due to the 
decrease in dimensionality. This is clear from Fig. 2, which shows the 2D rods of magnetic 
scattering co-existing with 3D C magnetic Bragg peaks for the x = 0.0125 sample, only at low 
temperature. In the HTS literature, the region of the phase diagram between 3D C AF order and a 
superconducting ground state, which is typically 0.02 $\leq$ x $\leq$ 0.05, is often refered to as 
a spin glass regime \cite{Birgeneau_JPhysSocJpn_2006}. This is correct in that the ground state 
spin correlations within the orthorhombic basal plane of these samples are finite and 
elastic. However, most importantly, the spin correlation lengths between orthorhombic planes have  gone to 
$\approx$ zero resulting in distinct rods of magnetic scattering; that is the layers are decoupled.  
This reduction in magnetic dimensionality from 3D to 2D, on its own, would be 
expected to strongly suppress any ordering transition in such a layered system, and indeed this is what 
is observed. It is an interesting observation that the C spin structure within the orthorhombic 
plane leads to a 3D structure while the IC spin structure within the orthorhomic plane displays 
a 2D ground state.

Finally, although it was first observed some time ago in $La_{2-x}Sr_{x}CuO_{4}$\cite{Yamada_PRB_1998, Wakimoto_PRB_2000} and more recently in $La_{2-x}Ba_{x}CuO_{4}$\cite{Dunsiger_PRB_2008}, 
it bears repeating that the quantum critical point between non-superconducting and superconducting ground states in $La_{2-x}Ba_{x}CuO_{4}$, near 
x $\sim$ 0.05, is coincident with the rotation in the 2D IC spin structure from diagonal to parallel. This also provides strong evidence for an intimate 
connection between the 2D magnetism and the superconducting properties in these HTS systems.

\begin{figure}
\centering
 \includegraphics[width=0.6\textwidth,height=!]{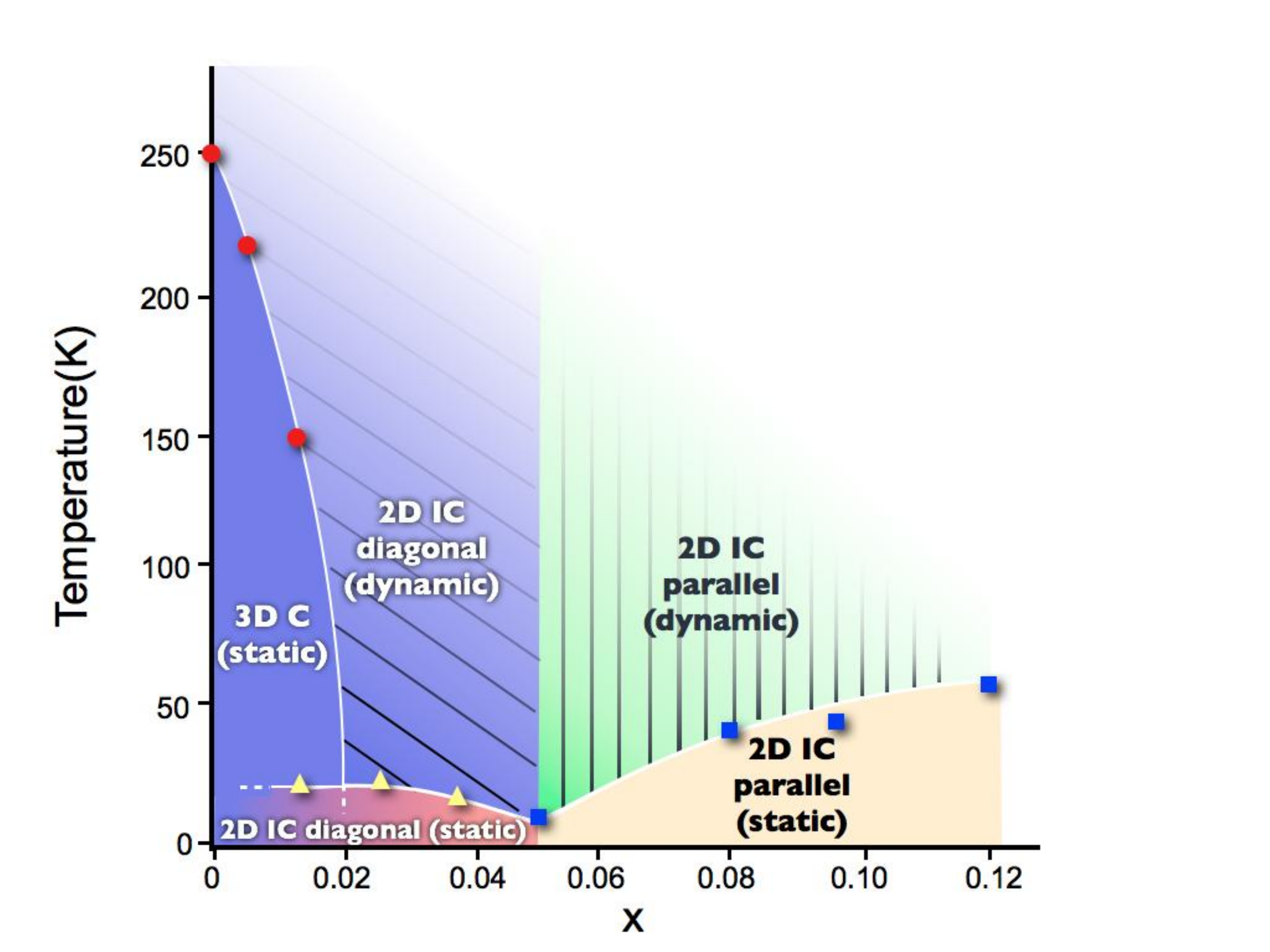} 
 \caption{Magnetic phase diagram for $La_{2-x}Ba_{x}CuO_{4}$ as determined by magnetic order parameter measurements on $La_{2-x}Ba_{x}CuO_{4}$ crystals with 
x $\leq$ 0.125. \cite{Dunsiger_PRB_2008,Hucker_PRB_2011}. 3D static C magnetic order gives way to 2D static (on the time scale of high energy 
resolution neutron measurements) diagonal IC order for x $\ge$ 0.02, with a co-existence between the two at low temperatures for smaller values of x. 
At an x $\sim$ 0.05 quantum critical point, the 2D IC ordering wavevector rotates from diagonal to parallel, relative to the pseudo-tetragonal axes, and this is 
coincident with the onset of a superconducting ground state\cite{Dunsiger_PRB_2008}. Dynamic 2D IC fluctuations persist to temperatures much higher than those 
characterizing the onset of static 2D order. These fade continuously with increasing temperature and inhabit much of the phase diagram associated 
with the ``pseudogap" phase.}
 \label{Phase_Diagram}
\end{figure}

\section{Conclusions}

We have carried out extensive neutron scattering measurements on the static and low energy, dynamic commensurate (C) and incommensurate (IC) magnetism in lightly-doped 
$La_{2-x}Ba_{x}CuO_{4}$ (LBCO). We have shown the two dimensional (2D) IC static order to be characterized by the appearance of rods of elastic, diagonal IC scattering with 
long but finite correlation lengths within the basal plane, and essentially zero correlation length along L. Moreover, below the 2D IC effective ordering temperature, T$_{N(2DIC}$, these rods are elastic on 
the energy scale of 0.1 meV, which is $\sim$ 1 K or less. We can understand the suppression of the 2D IC effective ordering temperature 
relative to the 3D C ordering temperatures displayed by nearby concentrations as a consequence of the reduction in magnetic dimensionality, rather than being due to proximity to a competing superconducting ground state.

A phase diagram based solely on magnetic order parameter measurements, and constructed using 3D C long range order as well as effective 2D IC static magnetic order transitions for all LBCO samples with $x \leq 0.125$ is shown to display much of the same complexity as that corresponding to standard phase diagrams relevant to high temperature superconductivity. This stresses the strong correlation between magnetism and the exotic charge correlation physics, including superconductivity itself, in this family of high temperature superconductors. Our measurements at low temperatures show a systematic suppression of the low energy dynamic susceptibility as a function of increasing doping within the lightly-doped regime $x \leq 0.035$, presaging the appearance of superconducting ground states for $x \geq 0.05$.

All samples studied in this paper, other than x = 0, display 2D diagonal IC static magnetism at low temperatures within their ground states. Interestingly, we find that the corresponding dynamic IC magnetism exists both at low 
temperatures as well as on a much higher temperature scale, comparable to nearby 3D C ordering temperatures. The temperature dependence of this dynamic IC magnetism does not change quickly with doping at these low dopings. This 2D dynamic IC magnetism inhabits much of the phase diagram associated with pseudo-gap physics, and there appears to be no characteristic transition temperature associated with these fluctuations; rather their temperature evolution is characteristic of crossover phenomnea.

\begin{acknowledgments}
We would like to acknowledge useful conversations had with A.J. Berlinsky, C. Kallin, G.M. Luke, 
J. P. Clancy, K. Fritsch, A. Dabkowski, and T. Timusk. We would also like to acknowledge M. D. Lumsden for technical assistance with the 
measurements on HB3. Research using ORNL's High Flux Isotope Reactor was sponsored by the Scientific User Facilities Division, 
Office of Basic Energy Sciences, U.S. Department of Energy. This work was supported by NSERC of Canada. 
This work utilized facilities supported in part by the National Science Foundation under Agreement No. DMR-0944772.
The identification of any commercial product or trade name does not imply endorsement or recommendation by the National Institute of Standards and Technology.
\end{acknowledgments}

\bibliographystyle{unsrt}
\bibliography{cuprate}

\end{document}